\documentclass[11pt]{article}
\usepackage{amsmath,amssymb}
\usepackage{epsfig}

\newtheorem{theorem}{Theorem}

\newcommand{\eat}[1]{}

\title {New model for rigorous analysis of LT-codes}
\author{Elitza N.\ Maneva and Amin Shokrollahi}

\begin{document}

\maketitle

\begin{center}
\begin{abstract}
We present a new model for LT codes which simplifies the analysis of
the error probability of decoding by belief propagation.
For any given degree distribution, we provide the first rigorous
expression for the limiting error probability as the length of the
code goes to infinity via recent results in random hypergraphs
\cite{dn}. For a code of finite length, we provide an algorithm for computing
the probability of error of the decoder. This algorithm improves the one of 
Karp, Luby, and Shokrollahi \cite{kls} by a linear factor.
%in time $O(n^2\log n)$ improving on the previous
%$O(n^3 \log^2 n)$ algorithm from \cite{kls}.
\end{abstract}
\end{center}

\section{Introduction}

Fountain codes were originally introduced in \cite{blmr} and were
designed for robust and scalable transmission of data over lossy
networks. Given a vector of input symbols $(x_1, x_2, \ldots, x_k)$, a
fountain code generates a stream of output symbols to be sent over the
network. Each output symbol is generated independently by sampling
from a fixed distribution on subsets of the input symbols and adding
the symbols in the chosen subset. The sequence of output symbols, together with
the positions of the input symbols whose sum they represent, is sent
over a lossy network. The input word is decoded using the belief
propagation algorithm which takes only linear time.  The probability 
that the belief propagation decoder fails depends on the
distribution from which output symbols were generated and on the
number $n$ of output symbols received.

Analysis of the error probability to date has been carried out under the 
assumption of a fixed
number of received output symbols $n$. Here we will change this
assumption and say that the number of output symbols received is a
random variable with mean $n$. This assumption makes sense in
applications and is not significantly different from the case of fixed
number of output symbols, because the random variable is
highly concentrated around $n$. We will define the exact
distribution in the following section. We refer to this as the Poisson
model because the number of output symbols approaches the Poisson
distribution as $k$ goes to infinity. Intuitively, the Poisson model
adds further independence between the random variables involved in the
error-probability calculations, and thus significantly simplifies the
analysis.

We will apply the new model to the analysis of a particular kind of
fountain codes - the LT codes introduced by Michael Luby in \cite{lt}.
The output symbols in LT codes are generated in the following way: $d$
is chosen from a fixed probability distribution $\Omega=(\Omega_1,
\Omega_2, \ldots \Omega_k)$ on the set ${1, 2, \ldots, k}$, after
which the parity of $d$ random input symbols is computed.

We are interested in two questions. Firstly, we look for an
analytic expression for the limiting error-probability of belief
propagation. The second question is that of designing an algorithm to compute
the error-probability for finite-length codes. 

The asymptotic analysis of LT codes to date has been based on a
heuristic calculation, using the fact that in the limit of $k$ going to
infinity the belief propagation iterations behave as if on a tree
graph. With the new model we can apply recent results in the analysis
of processes on random hypergraphs \cite{dn} to give an exact
expression for the portion of symbols that can be decoded by belief
propagation as $k$ goes to infinity.

For the finite-length analysis of LT codes, Karp, Luby, and Shokrollahi
\cite{kls} proposed a dynamic programming algorithm. The size of the 
table is $O(n^3)$ and each entry is computed using $O(n^2)$ of the
previous entries. Using generating polynomials representation and
fast multi-point evaluation and interpolation of polynomials, the
complexity of the algorithm is $O(n^3\log^2n)$.  The Poisson model
permits us to reduce the dynamic programming recursion to a table of
size $O(k^2)$, and each entry is computed from $O(k)$ of the previous
ones. Using generating polynomials representation, the complexity is
reduced to $O(k^2\log k)$.

In the next section we will review the factor graph representation of
LT codes, the belief propagation algorithm for them, and we will
define the new model precisely. Section \ref{sec:as} is dedicated to
the asymptotic analysis, and Section \ref{sec:fin} to the finite
length analysis. We will conclude with a brief discussion of open
problems.

\section{Background and Definitions}

\begin{figure}
\label{fig:factor-graph}
~\epsfig{file=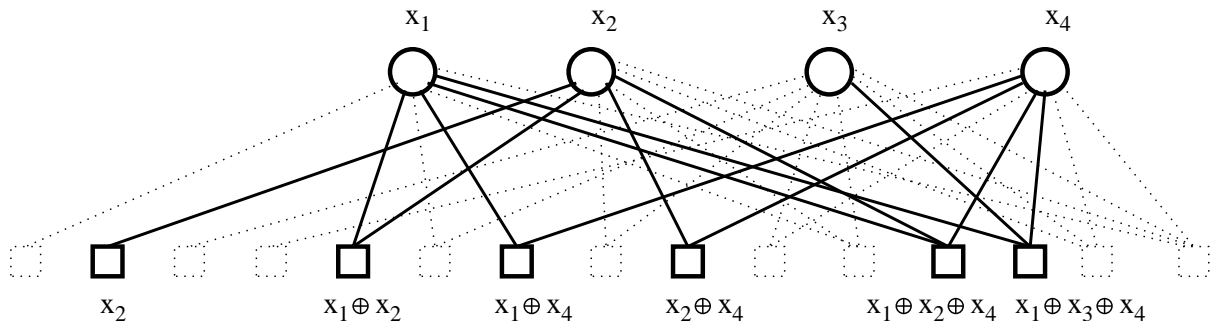, width=6.3in}
\caption{Round nodes denote input symbols and square nodes denote output 
symbols. Every subset of input symbol corresponds to a potential
output symbol, taking as value their sum mod 2. Each output symbol is
generated independently with probability that depends on the size of
the set. The solid square nodes correspond to output symbols that were
generated and received.}
\end{figure}

It is convenient to think of the set of input and output symbols as
the vertices of a bipartite graph. Every output symbols is connected
by an edge to all input symbols in the set whose sum it represents as in 
figure \ref{fig:factor-graph}.  

\subsection{Belief Propagation}

In the setting of fountain codes the belief propagation algorithm is
very simple.  If there is an output symbol with a single undecoded
neighbor, then the value of that input symbol can be computed. In this
case, we say that a {\it decodable} input symbol becomes {\it
uncovered} or {\it decoded}.  Uncovering one symbol may result in
other input symbols becoming decodable, and so on. The process stops
when there are no decodable input symbols, or equivalently, there are
no output symbols with a single undecoded neighbor. We refer to the
set of decodable input symbols as the \emph{ripple} (note that in
\cite{kls} the ripple is, instead, the set of output symbols that have
only one undecoded symbol).  At every step, one input symbol leaves
the ripple and 0 or more input symbols join the ripple.
%The process of belief propagation decoding is identical to the {\em hypergraph
% collapse} process of \cite{dn}.

\subsection{The Poisson model for LT codes}

For a given set of input symbols of size $d$, let $p_d =
{n\Omega_d}/{\binom{k}{d}}$ be the probability that an output symbol
representing the parity of this set was received.  
Then by linearity of expectation the expected number
of distinct output symbols is exactly $n$ and the expected number of
output symbols from sets of size $d$ is $n\Omega_d$.

Let $D$ be the largest degree with positive probability.  Let $N_d$
for $d=1, \ldots,D$ be a random variable denoting the total number of
output symbols of degree $d$, and $N=\sum_{d=1}^D N_d$ is the total
number of output symbols. The distribution of $N_d$ is binomial
$B(\binom{k}{d}, p_d)$. By concentration inequalities in \cite{janson}: 
\begin{eqnarray}
\Pr[N\ge n+\Delta] &\le&
\exp\Biggl(-\frac{\Delta^2}{2(n+\Delta/3)}\Biggr), \label{eqn1}
\\
\Pr[N\le n-\Delta] &\le& \exp\Biggl(-\frac{\Delta^2}{2n}\Biggr) .
\label{eqn2}
\end{eqnarray}

\section{Asymptotic Analysis of LT Codes}
\label{sec:as}

The above random model is almost identical to the Poisson random
hypergraph model of Darling and Norris \cite{dn}. The process that
they study, called the {\it hypergraph collapse process}, is identical
to the uncovering of input symbols in the belief propagation
algorithm. In order to restate their result in our setting, we need
some notation. Let $n=(1+\delta)k$ for some constant $\delta\ge0$. Let
$\beta_1=-\ln(1-(1+\delta)\Omega_1)$ and
$\beta_d=(1+\delta)\Omega_d$, for $d=2, \ldots, k$. From these we define
the power series 
$$\beta(t)=\sum_{d\ge 0}\beta_d t^d$$
and its derivative:
$$\beta'(t)=\sum_{d\ge 1}d \beta_d t^{d-1}.$$
The statement of the theorem is in terms of the roots of the function
$\beta'(t)+\log(1-t)$. Let
$$z^*=\inf\{t\in [0, 1): \beta'(t)+\log(1-t)<0 \} \wedge 1$$
and suppose there are no roots of $\beta'(t)+\log(1-t)$ in $[0,
z^*)$. Notice that in particular if $\beta$ is a polynomial (as is
the case in the LT-codes setting) then $z^*<1$.

\begin{theorem}{\bf \cite{dn}}
Assuming $z^*<1$ and there are no roots of $\beta'(t)+\log(1-t)$ in
$[0, z^*)$, then as $k$ goes to infinity the fraction of recoverable
input symbols goes to $z^*$ in probability.
\end{theorem}

Therefore as a first test for the quality of a particular degree
distribution, one can compute the roots of $\beta'(t)+\log(1-t)$. In
fact, $(1-t)(\beta'(t)+\log(1-t))$ is the expected fraction
of output symbols which have a unique undecoded neighbor, when
fraction $t$ of the input symbols have been decoded. This is equivalent
to the expression obtained from the tree analysis.

\section{Finite-length Analysis of LT Codes}
\label{sec:fin}

For codes of finite length, we are interested in calculating the
probability that all input symbols can be recovered. In this section
we give an algorithm for computing this probability for a given
degree distribution.

%\begin{theorem} 
%\end{theorem}

\subsection{Recursion of probability distributions}

Let $X_u$, for $u=1, \ldots, k$, be random variables that denote the
size of the ripple when $u$ symbols are undecoded, or equivalently,
we will sometimes say \emph{at step $u$}.  
In particular, $X_k$ is the number of input
symbols for which a degree-1 output symbol was generated.  This number
has a binomial distribution $B(k, p_1)$. The decoding process stops when
$X_u=0$.  The distribution of the size of the ripple at step $u-1$
depends only on the size of the ripple at step $u$. If
$X_u=0$ then the process stops and $X_{u-1}=0$. If $X_u>0$, then one
of the symbols in the ripple is decoded. This results in $Y_u$ input
symbols joining the ripple. $Y_u$ is distributed as the binomial
distribution $B(u-X_u, q_u)$, where $q_u$ is the probability that a
symbol joins the ripple at step $u$ (when the $(k-u+1)$-st symbol is
decoded).  An input symbol $a$ joins the ripple at this time if and
only if there is an output symbol with neighbors: the symbol $a$, the
last decoded symbol, and any set of symbols among the other $k-u$
decoded symbols. Therefore,
$$q_u=1-~\prod_{d=2}^{\min\{D, k-u+2\}}~(1-p_d)^{\binom{k-u}{d-2}}.$$ Finally,
$X_{u-1} = X_u - 1 + Y_u$. Therefore, for every $1 \le r \le u-1$ and
$1 \le s \le r+1$,
\begin{eqnarray*}
\Pr[X_{u-1}=r~|~ X_u=s] &=& \Pr[Y_u=r-s+1] 
= \binom{u-r}{r-s+1}~~q_u^{r-s+1}~(1-q_u)^{u-r-1},
\end{eqnarray*}
This gives us an expression for the distribution of $X_{u-1}$ in terms of 
the distribution of $X_u$:
\begin{eqnarray*}
\Pr[X_{u-1}=r] &=& \sum_{s=1}^{r+1} \Pr[X_u=s] \Pr[X_{u-1}=r | X_u=s].
\end{eqnarray*}

The probability that belief propagation cannot complete the decoding
is exactly the probability that $X_{1}=0$. This can be computed
by dynamic programming. Let $Q(u, r)$ be 
$\Pr[X_u=r]$, for every $u=0,\ldots, k$, and $r=0, \ldots, u$. Then
\begin{eqnarray*}
Q(k, r) &=& \binom{k}{r}~ p_1^r~ (1-p_1)^{k-r}, \mbox{ for } r=0,
\ldots, k\\ 
Q(u-1, r) &=& \sum_{s=1}^{r+1} Q(u, s)~ 
\binom{u-s}{r-s+1}~~q_u^{r-s+1}~(1-q_u)^{u-r-1}, 
\mbox{ for } r=1, \ldots, u-1\\
Q(u-1, 0) &=& Q(u, 0)+Q(u, 1)~(1-q_u)^{u-1}.
\end{eqnarray*}
Finally, the probability of error of the decoder is $Q(1, 0)$.

\subsection{Complexity of the algorithm}

To compute the values for $(1-q_u)$ for $u=1, \ldots, k$, we proceed
again by dynamic programming. We store the values for all of the
factors $f(u, d)=(1-p_d)^{\binom{k-u}{d-2}}$ for every $d=2,3, \ldots,
\min\{D, k-u+2\}$.  There are $O(Dk)$ entries and $f(u-1, d)=
f(u, d)^\frac{k-u+1}{k-u-d+3}$, which takes $O(\log k)$ operations to 
compute. Therefore precomputing $q_u$ takes $O(Dk\log k)$ operations.

The recursion for $Q(u, r)$ can be computed more efficiently if we
represent it by generating polynomials. We will proceed in a manner
similar to \cite{kls}. Let $Q_u(x)=\sum_{r=1}^{u}Q(u, r)x^{r-1}$.
Then the recursion can be written as:
\begin{eqnarray*}
Q_k(x) &=& \frac{1}{x}\bigl((p_1x+(1-p_1))^k-(1-p_1)^k\bigr),\\ \\ 
Q_{u-1}(x) &=&
\frac{(q_ux+(1-q_u))^{u-1}}{x}~~Q_u\Bigl(\frac{x}{q_ux+(1-q_u)}\Bigr) 
- \frac{(1-q_u)^{u-1}}{x}~~Q_u(0).
\end{eqnarray*}
Finally, the probability of success is $Q_1(x)$ (which is a constant). We
compute the sequence of polynomials $Q_u(x)$ for $u=k, \ldots, 1$, in
the following way: Suppose we have computed the coefficients of
$Q_u$. We choose $k$ non-zero points $\hat{x}_1, \hat{x}_2, \ldots
\hat{x}_k$, and compute 
$\hat{y}_i=\frac{\hat{x}_i}{q_u\hat{x}_i+(1-q_u)}$ for $i=1,
\ldots, k$. We evaluate $Q_u$ at the points $\hat{y}_i$ using the multipoint
evaluation algorithm, which takes $O(\log k)$ operations per
point. Given these values, evaluating $Q_{u-1}(\hat{x}_i)$ takes another
$O(\log k)$ operations. We can then interpolate the coefficients of
$Q_u$ by fast polynomial interpolation, which takes $O(k\log k)$
operations. Therefore, the time complexity of our algorithm is
$O(k^2\log k)$.

\subsection{Implications for the case with fixed number of output symbols}

The algorithm above outputs the probability that belief
propagation fails, given that the number of output symbols is a random
variable with expected value $n$, as described. 
Let's denote this probability by $P_p(n)$.
We can use this algorithm to get bounds for the probability that belief
propagation fails, when there is a fixed number of output symbols,
which we denote by $P_f(n)$. We use the fact that the probability
that the decoder fails is monotone decreasing in the number of output
symbols, and
\begin{eqnarray*}
P_p(n) = \sum_{\hat{n}=0}^{2^k-1} \Pr[N=\hat{n}] \times P_f(\hat{n}).
\end{eqnarray*}
Let $n_1\le n \le n_2$. Using the concentration inequalities
(\ref{eqn1}) and (\ref{eqn2}) we get the bounds:
$$P_p(n_1)-\exp\Bigl(-\frac{3(n-n_1)^2}{2(2n_1+n))}\Bigr) ~\le~ P_f(n)
~\le~ \frac{P_p(n_2)}{1-\exp\Bigl(-\frac{(n_2-n)^2}{2n_2}\Bigr)}.$$

\section{Discussion}

Our approach presented here is applicable to general fountain codes,
as well as some classes of LDPC codes. Di et al. \cite{finLDPC} gave
algorithms for the finite-length analysis of regular LDPC codes
(i.e. left and right degrees are constant). Our method is applicable
to codes with Poissonian degree distribution on the right.

% For final version
%\section{Acknowledgements}
%This work was done while EM was visiting the Algo group at EPFL, and
%she would like to acknowledge their hospitality. Thanks to Alistair
%Sinclair for helpful discussions. The first author was funded by grant
%NSF:?.

\end{document}